# Who Followed the Blueprint?
# Analyzing the Responses of U.S. Federal Agencies to the Blueprint for an AI Bill of Rights

Darren Lage[1], Riley Pruitt[2], and Jason Ross Arnold[1]

[1] Department of Political Science, Virginia Commonwealth University
[2] Department of Computer Science, Virginia Commonwealth University

## Abstract

This study examines the extent to which U.S. federal agencies responded to and implemented the principles outlined in the White House's October 2022 "Blueprint for an AI Bill of Rights." The Blueprint provided a framework for the ethical governance of artificial intelligence systems, organized around five core principles: safety and effectiveness, protection against algorithmic discrimination, data privacy, notice and explanation about AI systems, and human alternatives and fallback.

Through an analysis of publicly available records across 15 federal departments, the authors found limited evidence that the Blueprint directly influenced agency actions after its release. Only five departments explicitly mentioned the Blueprint, while 12 took steps aligned with one or more of its principles. However, much of this work appeared to have precedents predating the Blueprint or motivations disconnected from it, such as compliance with prior executive orders on trustworthy AI. Departments' activities often emphasized priorities like safety, accountability and transparency that overlapped with Blueprint principles, but did not necessarily stem from it.

The authors conclude that the non-binding Blueprint seems to have had minimal impact on shaping the U.S. government's approach to ethical AI governance in its first year. Factors like public concerns after high-profile AI releases and obligations to follow direct executive orders likely carried more influence over federal agencies. More rigorous study would be needed to definitively assess the Blueprint's effects within the federal bureaucracy and broader society.

# Introduction

The rapid development and implementation of machine learning in health care, law enforcement, finance, science, and other domains have generated critical questions about its impact on individuals and society. In October 2022, the White House responded with the "Blueprint for an AI Bill of Rights," a non-binding framework outlining principles to "guide the design, use, and deployment of automated systems" [1][2]. Technically a white paper published by the White House's Office of Science and Technology Policy (OSTP), the Blueprint set the stage for a structured and principled approach to ethical AI governance in the U.S. It pre-dated President Biden's Executive Order on the Safe, Secure, and Trustworthy Development and Use of Artificial Intelligence by a year [3].

The Blueprint organizes its guidance around five core principles. AI systems should be (1) "safe and effective." Citizens should enjoy (2) "algorithmic discrimination" and (3) "data privacy" protections, and be given (4) "notice and explanation" of an automated system's presence in an interaction, as well as an understanding about "how and why it contributes to outcomes" that affect them. Finally, citizens interacting with systems should have the opportunity to "opt out, where appropriate, and have access to a person who can quickly consider and remedy problems," ensuring (5) "human alternatives, consideration, and fallback."

While the Blueprint is "non-binding and does not constitute U.S. government policy," the OSTP makes clear that it "advance[s] President Biden's vision," specifically his focus on "root[ing] out inequity, embed[ding] fairness in decision-making processes, and affirmatively advanc[ing] civil rights, equal opportunity, and racial justice in America," as articulated in the Executive Order (13985) he issued on his first day in office [4]. While it is not specifically targeted toward U.S. federal departments or agencies, the document makes clear that it "applies to... automated systems that... have the potential to meaningfully impact the American public's rights, opportunities, or access to critical resources or services" (p. 8). No other entity in American life has a greater impact on its citizens than the U.S. government.

In October 2023, a year after the Blueprint's publication, we began to examine the extent to which Biden's agencies heeded its call.[1] In this paper, we compare whether and how executive branch departments have engaged with the Blueprint, through an analysis of their publicly accessible records. Did Biden's departments follow his lead?

---

[1] Alex Engler examined changes across the government in November 2022, a month after the Blueprint's publication. See Alex Engler, "The AI Bill of Rights Makes Uneven Progress on Algorithmic Protections," *Lawfare*, November 21, 2022, available at https://www.brookings.edu/articles/the-ai-bill-of-rights-makes-uneven-progress-on-algorithmic-protections/.

# Findings

Our exploratory study suggests the answer is no. Only five of fifteen U.S. federal departments – Commerce, Labor, State, Treasury, and Veterans Affairs – mentioned the Blueprint by name in any public document, which is one indication of its limited influence. While twelve departments, including the five, did work aligned with at least one of the five core principles after October 2022, we could not rule out the possibility that all or some of those efforts had no connection with the Blueprint.[2]

Moreover, several departments had already initiated work related to Blueprint principles before its release. For example, the Department of Health and Human Services (HHS) in January 2021 published a broad Artificial Intelligence (AI) Strategy, partly as a response to Executive Order 13960's (December 2020) emphasis on "trustworthy AI" and Executive Order 13859's (February 2019) call to "maintain American leadership in AI" (p. 3) [5][6][7]. In September 2021, HHS's Office of the Chief Artificial Intelligence Officer (CAIO) (created in December 2020) followed up by releasing the Trustworthy AI Playbook, which asks operating and staff divisions to apply six principles "across all phases of an AI project": fair/impartial, transparent/explainable, responsible/accountable, robust/reliable, privacy-preserving, and safe/secure [8][9][10]. Throughout 2022, HHS's commitment continued through internal briefings (e.g., on algorithmic bias), training sessions, and, after October, adherence to the Blueprint.[3]

The Defense Department, in July 2018, announced a plan to "establish a set of AI Principles for Defense" [11].[4] Fifteen months later, the Defense Innovation Board put forward a department-wide framework for the ethical use of AI emphasizing five core principles: responsibility, reliability, governability, traceability, and equity [12]. In December 2020 (days before E.O. 13960), the Homeland Security Department released an "Artificial Intelligence Strategy," pledged to "ensure… that DHS Components have measures in place to increase transparency, accountability, and to regularly monitor AI systems for potential bias and error" [13] The Commerce Department's National Institute of Standards and Technology (NIST), which had been establishing benchmarks and evaluation standards for the industry since the 1990s, responded to the deep learning revolution with "Towards a Standard for

---

[2] The exceptions are Interior and Transportation.

[3] Executive Order 14110 (October 2023) tasked HHS with creating an AI Task Force to advance the "responsible deployment and use of AI and AI-enabled technologies in the health and human services sector."

[4] This came partly in response to a protest by Google employees who objected to a partnership with DOD on Project Maven. For a good overview, see Paul Scharre, *Four Battlegrounds: Power in the Age of Artificial Intelligence* (New York: W.W. Norton, 2023), pp. 52-9. A month earlier, DOD established the Joint Artificial Intelligence Center (JAIC), which it later (February 2022) merged into the Chief Digital and Artificial Intelligence Office (CDAO).

Identifying and Managing Bias in Artificial Intelligence" (2020), "Trust and Artificial Intelligence" (2020), and "Four Principles of Explainable Artificial Intelligence" (2021), which emphasized principles found in the Blueprint [14][15][16]. NIST also launched an ambitious AI Risk Management Framework initiative in July 2021, a "consensus-driven, open, transparent, and collaborative process" that placed several key "characteristics of trustworthy AI systems" at the forefront, including validity and reliability, accountability and transparency, safety, security and resiliency, privacy, and fairness ("with harmful bias managed")(p. 12) [17][18].[5] The Energy Department similarly released an AI Risk Management Playbook in August 2022 "to support responsible and trustworthy… AI use and development" [19].

After the Blueprint's release, several departments took steps that aligned with its framework. For example, in April 2023, the Justice Department's Civil Rights Division, with the Consumer Financial Protection Bureau, Equal Employment Opportunity Commission, and Federal Trade Commission, released a "Joint Statement on Enforcement Efforts Against Discrimination and Bias in Automated Systems" [20]. The Labor Department's Office of Disability Employment Policy, with the public-private Partnership on Employment and Accessible Technology, held a four-hour online meeting of public sector workers, technologists, and representatives of civil rights and disability organizations to discuss the future of AI in hiring practices, with a focus on minimizing the risks of discrimination [21]. In May 2023, the Education Department's Office of Educational Technology published "Artificial Intelligence and the Future of Teaching and Learning: Insights and Recommendations," which emphasizes several core principles aligned with the Blueprint, including: humans-in-the-loop; privacy and data security; transparent, accountable, and responsible use; minimization of bias and the promotion of fairness; and explainability [22].

However, we found it difficult to connect any of these steps with the Blueprint. Other motivating factors, including the need to comply with executive orders, and an incentive to respond to increased public awareness and concerns after OpenAI released ChatGPT in November 2022, probably had a greater impact on departmental decisions to act. Sometimes these causal factors were clearly invoked, as with the Treasury Department's appointment of a Responsible AI Official in December 2022 as part of its "Executive Order 13960

---

[5] A third and final draft of the AI RMF was released on January 23rd, 2023. Other examples show departments touching upon Blueprint-like principles, but not directly addressing them in ways that would structure their work going forward. The Justice Department sued Meta for "discriminatory advertising in violation of the Fair Housing Act (FHA)," which it argued resulted from algorithmic bias. The case, settled in June 2022, addressed a key Blueprint principle but did not reflect a Department-wide effort against bias. The Department of Transportation emphasized safety and, to a lesser extent, transparency and accountability, in its work with AI before October 2022. But, instead of focusing on the responsible use of AI, it utilized AI to increase safety (etc.) in existing and developing programs. Similarly, the Department of Housing and Urban Development (HUD) used AI to reduce human bias in home lending, but did not (as far as we can tell) focus on algorithmic bias itself.

Consistency Plan" [23]. Health and Human Services AI Strategy, as noted earlier, came in part as a response to the same directive.

Moreover, department efforts made after October 2022 had precedents before the Blueprint. The Labor Department, as noted earlier, held a post-Blueprint event about hiring and workplace risks from AI. But its Equal Opportunity Commission hosted a similar discussion about AI risks to hiring and workplace equity in September 2022 [24]. Homeland Security, which had made strides before the Blueprint, continued its work with Directive 026-11 in September 2023, which authorized AI use in investigations with the requirement that officials follow "safeguards for privacy, civil rights, and civil liberties" [25]. The Commerce Department's NIST released its final version of the AI Risk Management Framework in January 2023 [18].

## Conclusion

Overall, the Blueprint seems to have had a limited to non-existent impact across federal departments. Although we found significant variation across units in types and levels of activity, the vast majority of departments appear to have made some sort of effort on Blueprint-related goals; Transportation appears to be the lone laggard here. However, whether any of those actions were connected to the Blueprint remained unclear. This lack of responsiveness to the White House document was not surprising, considering its non-binding and public-facing nature. Departments have a need to directly and explicitly respond to executive orders. Guidance from a high-profile but advisory office in the White House might have an influence on the bureaucracy, but those who follow it have no reasons to acknowledge it in official records or actions.

The Blueprint's influence might still be all over post-October 2022 actions, but identifying it would require a more rigorous, systematic study involving extensive interviewing and content analysis of all officials' statements in speeches, media appearances, and anywhere else they appear. It would also need a more systematic review of departments' sub-units, along with independent agencies and other bodies that exist outside of the departmental structure. Plus, considering the Blueprint's other audiences, such a study would need to research its impact in the private sector, and across American society more broadly (e.g., in higher education). However, our research suggests that work done in the name of the Blueprint is sparse. Moreover, the increased attention on AI policymaking in the months following its release has resulted in a flurry of similarly-aligned governance documents such as E.O. 14110, further drawing attention away from the Blueprint.

## Acknowledgments

This project was supported by the VCU Transformative Learning Fund. We thank Milos Manic, as well as Nate Eldering, Laila Leak, and Bindi Patel, all part of our AI Governance VIP (Vertically Integrated Project) team in 2023.

"Joint Statement on Enforcement Efforts against Discrimination and Bias in Automated Systems," https://www.justice.gov/crt/page/file/1581491/dl?inline.

[21] U.S. Department of Labor Office of Disability Employment Policy, "Department Of Labor Gathered Experts, Stakeholders to Ensure More Inclusive Hiring as Automated Technology Affects Decision-Making," https://www.dol.gov/newsroom/releases/odep/odep20230417.

[22] U.S. Department of Education Office of Educational Technology, "Artificial Intelligence and the Future of Teaching and Learning," May 2023, https://tech.ed.gov/ai-future-of-teaching-and-learning/.

[23] U.S. Department of the Treasury, "Executive Order 13960 Consistency Plan," December 2022, https://home.treasury.gov/system/files/136/Treasury-EO13960-Consistency-Plan.pdf.

[24] U.S. Equal Employment Opportunity Commission, "Decoded: Can Technology Advance Equitable Recruiting and Hiring? Opening statements," https://www.youtube.com/watch?v=6tVOccqVs0w.

[25] U.S. Department of Homeland Security, "Use of Face Recognition and Face Capture Technologies," DHS Directive 026-11, September 11, 2023, https://www.dhs.gov/sites/default/files/2023-09/23_0913_mgmt_026-11-use-face-recognition-face-capture-technologies.pdf.
7